# Output Width Signal Control In Asynchronous Digital Systems Using External Clock Signal

**Assist.Prof. Mihai Timiş,**
"Gh. Asachi" Technical University, Iaşi, Romania
Faculty of Automation and Computers

**ABSTRACT.** In present paper, I propose a method for resolving the timing delays for output signals from an asynchronous sequential system. It will be used an example of an asynchronous sequential system that will set up an output signal when an input signal will be set up. The width of the output signal depends on the input signal width, and in this case it is very short. There are many synthesis methods, like using a RC group system, a monostabil system in design of the asynchronous digital system or using an exetrnal clock signal, CK. In this paper will be used an exetrnal clock signal, CK.

## 1. Introduction

I will porpose an asynchronous digital system who will generate an output signal, named Z, drive by an logic signal named H (high), figure 1.
　　The fluence state graph is ilustrated in figure 2.
　　Based on figure 1,2 will be deducted the fluence state table, transition/output matrix, figure 3.

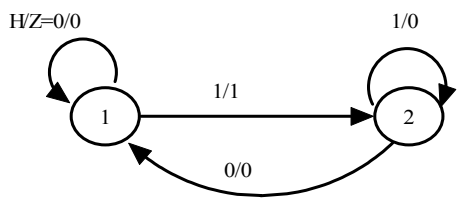

***Fig. 1***

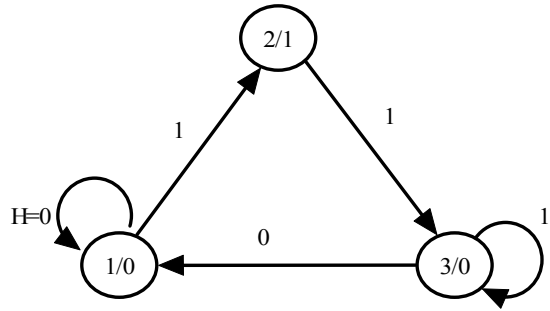

*Fig. 2*

| H  (y₁y₀) | $(y_1y_0)_{n+1}$ | | $z_n$ | |
|---|---|---|---|---|
|  | 0 | 1 | 0 | 1 |
| 0 0 | 0 0 | 0 1 | 0 | 0 |
| 0 1 | - - | 1 0 | 1 | 1 |
| 1 1 | - - | - - | - | - |
| 1 0 | 0 0 | 1 0 | 0 | 0 |

*Fig. 3*

The asynchronous digital system equations are:

$$\begin{cases} y_{1,n+1} = (y_0 + y_1 H)_n \\ y_{0,n+1} = (\overline{y_1} \overline{y_0} \overline{H})_n \\ z_n = (\overline{y}_1 y_0)_n \end{cases}$$

## 2. Method Description

Using an external clock signal CK with ½. filling range, will be generated only when the Z output signal will be triggered. The input signals are H, CK and output signal Z will be generated on CK high level. The state fluence table is illustrated on figure 4.

238

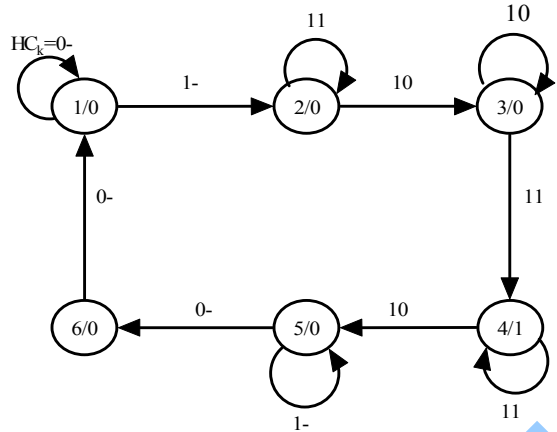

*Fig. 4*

The (2) state of the system graph is used for synchronisation of the Z output signal with the input CK clock signal.

The (6) state of the system graph is used for adjacent states transfers.

Fluence state table is ilustrated in figure 5.a, 5.b.

The asynchronous digital system equations are:

$$y_{2,n+1} = y_1(y_2 + \overline{y_0 c_k}) = \overline{\overline{y_1} + \overline{y_2 + \overline{y_0} + c_k}}$$

$$y_{1,n+1} = H(y_1 + y_0 \overline{c_k}) = \overline{\overline{H} + \overline{y_1 + \overline{y_0} + c_k}}$$

$$y_{0,n+1} = H(\overline{y_1} + y_0 \overline{c_k}) = \overline{\overline{H} + \overline{\overline{y_1} + \overline{y_0} + c_k}}$$

$$z_n = \overline{y_2} \overline{y_1} \overline{y_0} = \overline{y_0 + \overline{y_1} + y_2}$$

| Q/Z | HC$_k$ | | | | Cod stare |
|---|---|---|---|---|---|
| | 00 | 01 | 11 | 10 | y$_2$y$_1$y$_0$ |
| 1/0 | 1 | 1 | 2 | 2 | 0 0 0 |
| 2/0 | - | - | 2 | 3 | 0 0 1 |
| 3/0 | - | - | 4 | 3 | 0 1 1 |
| 4/0 | - | - | 4 | 5 | 0 1 0 |
| 5/0 | 6 | 6 | 5 | 5 | 1 1 0 |
| 6/0 | 1 | 1 | - | - | 1 0 0 |

*Fig. 5a: Fluence table*

| $HC_k$ $\diagdown$ $y_2y_1y_0$ | $(y_2y_1y_0)_n$ | | | | $z_n$ | | | |
|---|---|---|---|---|---|---|---|---|
| | 00 | 01 | 11 | 10 | 00 | 01 | 11 | 10 |
| 000 | 000 | 000 | 001 | 001 | 0 | 0 | 0 | 0 |
| 001 | --- | --- | 001 | 011 | 0 | 0 | 0 | 0 |
| 011 | --- | --- | 010 | 011 | 0 | 0 | 0 | 0 |
| 010 | --- | --- | 010 | 110 | 1 | 1 | 1 | 1 |
| 110 | 100 | 100 | 110 | 110 | 0 | 0 | 0 | 0 |
| 111 | --- | --- | --- | --- | - | - | - | - |
| 101 | --- | --- | --- | --- | - | - | - | - |
| 100 | 000 | 000 | --- | --- | 0 | 0 | 0 | 0 |

***Fig. 5b***

The design of the digital system implemented with MMC– 4002 circuits, is illustrated in figure 6.

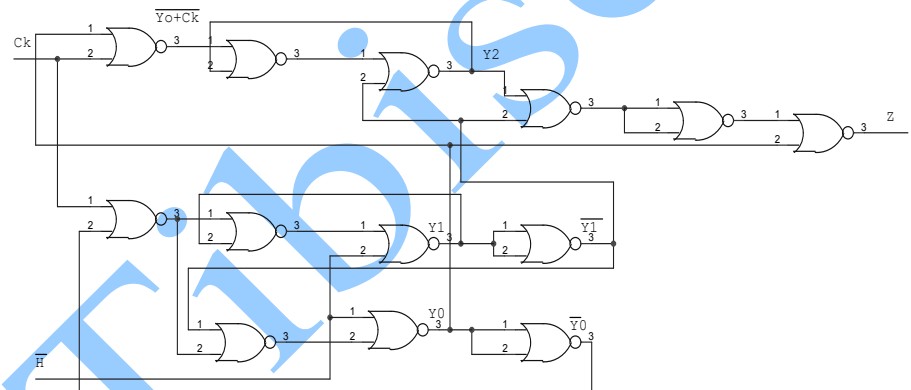

***Fig. 6***

The state graph table from figure 4 can be reduced like in figure 7.

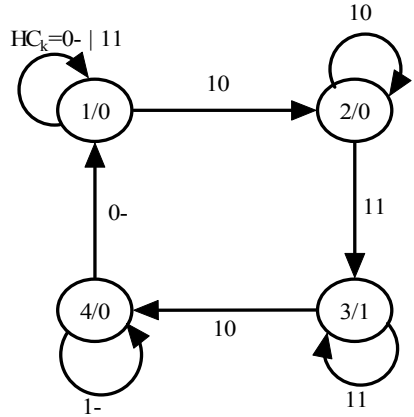

*Fig. 7*

The asynchronous digital system reduced equations are:

$$y_{1,n+1} = y_0 c_k + y_1 H = \overline{\overline{\overline{y_0} + \overline{c_k}} + \overline{\overline{y_1} + \overline{H}}}$$

$$y_{0,n+1} = y_0 c_k + \overline{y_1} H \overline{c_k} = \overline{\overline{\overline{y_0} + \overline{c_k}} + \overline{y_1 + \overline{H} + c_k}}$$

$$z_n = y_1 y_0 = \overline{\overline{y_1} + \overline{y_0}}$$

Based on upper equations, the design of the digital system implemented with MMC– 4002 circuits, is illustrated in figure 8.

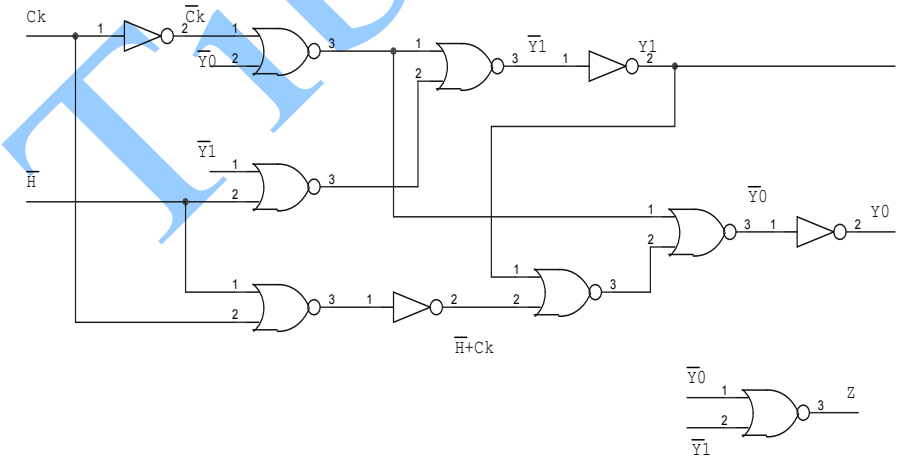

*Fig. 8*

## Conclusions

We will propose two methods for enlarge the width of the Z output signal:
- Using a RC circuit (the Z output signal will drive a monostabil who will generate a signal, proportional with the RC constant).
- Using a monostabil circuit (the Z output signal will drive a monostabil who will generate a M signal, proportional with the RC constant).